\pdfoutput=1
\documentclass[preprint,pre,floats,showkeys,aps,amsmath,amssymb,superscriptaddress,10pt,nofootinbib]{revtex4-1}
\usepackage{graphicx}
\usepackage{natbib} 
\setcitestyle{square,comma,numbers,sort&compress} 
\usepackage{amsmath,amsfonts,amssymb}
\usepackage{array}
\usepackage{tabularx} 
\usepackage[left = 1.875in, top=1.875in,right=1.875in,bottom=1.875in,a4paper]{geometry}
\usepackage[dvipsnames]{xcolor} 
\usepackage{colortbl} 
\usepackage{comment} 
\usepackage{siunitx} 
\usepackage[caption=false]{subfig} 
\usepackage{tikz} 
\usepackage[compat=1.1.0]{tikz-feynman} 
\usepackage{cancel} 
\usepackage{hyperref}

\newcommand{\be}{\begin{equation}}
\newcommand{\ee}{\end{equation}}
\newcommand{\ben}{\begin{equation*}}
\newcommand{\een}{\end{equation*}}

\def\bea{\begin{eqnarray}}
\def\eea{\end{eqnarray}}
\def\bean{\begin{eqnarray*}}
\def\eean{\end{eqnarray*}}

\def\l2{\log_2\,}
\newcommand{\barr}{\begin{array}}
\newcommand{\earr}{\end{array}}

\newcommand{\bed}{\begin{displaymath}}
\newcommand{\eed}{\end{displaymath}}
\newcommand{\bal}{\begin{array}{ll}}
\newcommand{\eal}{\end{array}}

\def\ket#1{\vert\,#1\,\rangle}

\def\mc#1{\mathcal#1}

\def\m#1{\mathcal#1}

\newcolumntype{Y}{>{\centering\arraybackslash}X} 
\newcommand{\g}[1]{\mathbf{#1}} 
\newcommand{\vev}[1]{\langle #1 \rangle} 
\newcommand{\gb}[1]{\bar{\mathbf{#1}}} 
 
\newcommand{\matc}[1]{  \left(
\begin{array}{c}
 #1
\end{array}
\right)
} 
\newcommand{\rd}[1]{\textcolor{black}{#1}}
\newcommand{\red}[1]{\textcolor{black}{#1}}
\newcommand{\blue}[1]{\textcolor{black}{#1}}
\newcommand{\sky}[1]{\textcolor{black}{#1}}
\newcommand{\blu}[1]{\textcolor{black}{#1}}
\newcommand{\myboxed}[1]{%
  \rlap{\hspace*{\dimexpr\fboxrule+\fboxsep\relax}%
    \phantom{\m@th$\displaystyle#1$}}%
    \smash{\boxed{#1}}}

\makeatletter
\renewcommand{\p@subsection}{}
\renewcommand{\p@subsubsection}{}
\makeatother

\begin{document}

\title{\Large Stitching an Asymmetric Texture with $\mc T_{13} \times \mathcal{Z}_5$ Family Symmetry \\\vspace*{1cm}}

\author{M. Jay P\'erez} 
\email[Email: ]{mperez75@valenciacollege.edu}
\affiliation{\small{Valencia College, Osceola Science Department, Kissimmee, FL 34744, USA}}

\author{Moinul Hossain Rahat} 
\email[Email: ]{mrahat@ufl.edu}
\affiliation{\small{Institute for Fundamental Theory, Department of Physics,
University of Florida, Gainesville, FL 32611, USA }}

\author{Pierre Ramond} 
\email[Email: ]{ramond@phys.ufl.edu}
\affiliation{\small{Institute for Fundamental Theory, Department of Physics,
University of Florida, Gainesville, FL 32611, USA }}

\author{Alexander J. Stuart} 
\email[Email: ]{astuart@ucol.mx}
\affiliation{\small{Facultad de Ciencias-CUICBAS, Universidad de Colima, C.P. 28045, Colima, M\'exico}\\\vspace*{1cm}}

\author{Bin Xu } 
\email[Email: ]{binxu@ufl.edu}
\affiliation{\small{Institute for Fundamental Theory, Department of Physics,
University of Florida, Gainesville, FL 32611, USA }}



\begin{abstract}
\vskip 0.5cm
We propose $\mc T_{13} = \mc Z_{13} \rtimes \mc Z_3$ as the underlying non-Abelian discrete family symmetry of the asymmetric texture presented in \cite{rrx2018asymmetric}. Its mod $13$ arithmetic distinguishes each Yukawa matrix element of the texture. We construct a model of effective interactions that singles out the asymmetry and equates, without fine-tuning, the products of down-quark and charged-lepton masses at a GUT-like scale. 
\end{abstract}

\maketitle

\section{Introduction}

The observable Pontecorvo-Maki-Nakagawa-Sakata (PMNS) neutrino mixing matrix  is the overlap of the unitary matrix $\mc U^{(-1)}$ that mixes the charged-lepton Yukawa matrix $Y^{(-1)}_{}$  and a quasiunitary matrix $\m U^{}_{\mathrm{Seesaw}}$ that diagonalizes the $3\times 3$ Majorana matrix of the light neutrinos, $\mc M^{(0)}$, i.e., 

\be
\m U^{}_{\mathrm{PMNS}}={\m U^{(-1)}}^{\dagger}\,\m U^{}_{\mathrm{Seesaw}}.
\ee

\vskip 0.3cm
\noindent Thus, the observable mixing angles have two different origins: $\mc U^{(-1)}$ comes from $\Delta I_w = \frac{1}{2}$ electroweak physics, whereas in the seesaw mechanism, $\mc U_{\mathrm{Seesaw}}$ comes from unknown $\Delta I_w = 0$ physics; the PMNS matrix bridges the $\Delta I_w = \frac{1}{2}$ and $\Delta I_w = 0$ sectors. Two out of its three angles are large, with a much smaller  third ``reactor angle".  In contrast, the largest of the quark mixing angles is the Cabibbo angle. 

\vskip .3cm
In the $SU(5)$ extension of the Standard Model, the down-quark Yukawa matrix $Y^{(-{1\over 3})}_{}$  is  similar to the transpose of the charged-lepton Yukawa matrix $Y^{(-1)}$,
 
 \be 
 Y^{(-{1\over 3})}_{}\sim Y^{(-1)T}_{},
 \ee
implying that the left-handed charged-lepton unitary matrix $\mc U^{(-1)}$ is  \textcolor{black}{similar to} the right-handed down-quark unitary matrix $\mc V^{(-\frac{1}{3})}$. 

In a basis where the up-quark Yukawa matrix $Y^{({2\over 3})}_{}$ is diagonal, the left-handed unitary matrix of  $Y^{(-{1\over 3})}_{}$ is the Cabibbo-Kobayashi-Maskawa (CKM) matrix which contains only small angles. In $SU(5)$, a \emph{symmetric} down-quark Yukawa matrix leads to small left-handed mixings of the charged leptons. Its contribution provides a small ``Cabibbo haze" \cite{cabibbohaze, *everett2006viewing, *Everett:2006fq, *kile2014majorana}  to the angles of the seesaw mixing matrix. 

\vskip 0.2 cm

Before the value of the reactor angle $\theta_{13}$ was measured \cite{dayabay, *doublechooz, *reno}, the large atmospheric and solar mixing angles were approximately expressed by ``platonic'' mixing matrices, e.g., tribimaximal (TBM) \cite{harrison2002tri, *tbm2, *xing2002nearly, *he2003some, *wolfenstein1978oscillations}, bimaximal (BM) \cite{vissani1997study, *barger1998bi, *baltz1998solar, *georgi2000neutrinos, *stancu1999flatness}, and golden ratio mixings GR1 \cite{goldenratio, *everett2009icosahedral} and GR2 \cite{rodejohann2009unified, *adulpravitchai2009golden}.  All possess a maximal atmospheric angle and a vanishing reactor angle, differing in their prediction for the solar mixing angle.  When corrected via contributions from flavor-symmetric Yukawa matrices, the reactor angle expectations hovered around $4^{\circ}$-$5^{\circ}$ \cite{kile}, \blue{much less than its measured value}.  

\vskip 0.3 cm
\blue{These simple and beautiful mixing matrices may be salvaged if the Yukawa matrices are asymmetric \cite{chamoun2019phase, *alcaide2018fitting, *meroni2012supersymmetric, *RIVERAAGUDELO201989, *rachlin2018gauged, *petcov2018assessing, *Lu:2019gqp, *dicus2011generalized}. However, models based on an underlying family symmetry, where the $SU(5)$ quintets and decuplets transform as the \emph{same} representations of the group, can only single out symmetric and antisymmetric Yukawa matrices. This leads to two questions: (a) what asymmetry is required by the Yukawas to satisfy the experimental constraints; and, (b) which family symmetry group can naturally produce an asymmetry? }

\vskip 0.2cm
 \blue{A minimalist answer to the first question was provided by three of us  in a phenomenological texture with an asymmetry present in \emph{only} the $(31)$ element of $Y^{(-{1\over 3})}_{}$ and $(13)$ element of $Y^{(-{1})}_{}$   \cite{rrx2018asymmetric}.} It reproduces features of the quarks and charged leptons such as the CKM matrix, the Gatto-Sartori-Tonin (GST) relation \cite{gatto}, and the mass ratios between down quarks and charged leptons in the deep ultraviolet. The charged-lepton mixings are now of the order of the Cabibbo angle, so that when folded in with the unperturbed TBM mixing, they yield a reactor angle larger than its experimental value. 

The addition of a $ \cancel {CP}$ phase \cite{parida2019high, *Shimizu2019,  *ballett2014testing, *ballett2014testingatm, *antusch2018predicting, *king2019theory, *ahriche2018mono, *delgadillo2018predictions, *girardi2016leptonic, *ding2019status, *Chen:2019egu,  *liu2019further, *petcov2018discrete, *girardi2015predictions, *girardi2015determining, *girardi2016predictions, *dinh2017revisiting, *penedo2018low, *agarwalla2018addressing, *petcov2016theory, *ge2011z2, *ge2012residual, *Ding:2019zhn, *CarcamoHernandez:2018djj, *Chen:2015siy, *Chen:2018eou}  to the TBM matrix is necessary to lower $\theta_{13}$ to its Particle Data Group (PDG) value  \cite{tanabashi2018review}. \blue{This \emph{single} parameter brings the other two angles within $1\sigma$ of their PDG fit and predicts the $\cancel{CP}$ Jarlskog-Greenberg invariant \cite{jarlskog1, *greenberg1}, $|\mc J|\approx 0.028$, which matches with the central PDG value.}  


\vskip .2cm

\blue{In this work we propose an answer to the second question with a family symmetry  (see \cite{king2013neutrino, *tanimoto2015neutrinos, *meloni2017gut, *petcov2017discrete} and the references therein) based on the discrete group $\mc T_{13} = \mc Z_{13} \rtimes \mc Z_3$ \cite{ding2011tri, *hartmann2011neutrino, *kajiyama2011t13, *PhysRevD.85.013012}.  It explains the asymmetric term of the texture and yields the equality of the determinants of the matrices $Y^{(-{1\over 3})}_{}$ and $Y^{(-1)}_{} $, conforming to the down-quark to charged-lepton mass ratios at the GUT scale. }\rd{$\mc T_{13}$, however, allows some operators which spoil these features. Such operators can be naturally avoided and the determinant condition can be established successfully only when the family symmetry is extended to include a $\mc Z_5$ factor. }

\rd{It is useful to comment here that a complete flavor model would construct all Yukawa matrices of the Standard Model, i.e.,  $Y^{(2/3)},\,Y^{(-1/3)},\,Y^{(-1)}$, as well as generate a light neutrino mass matrix $\mc M^{(0)}$. As a first step in this direction, in this work we focus solely on the asymmetric matrices for the down quarks and charged leptons \cite{rrx2018asymmetric}, and show how they can naturally arise from the discrete family symmetry $\mc T_{13}\times \m Z_5$.  }

\rd{The asymmetric texture of \cite{rrx2018asymmetric} requires $Y^{(2/3)}$ to be diagonal, which, as we will show below, is natural to obtain with $\mc T_{13}\times \m Z_5$.  It also requires that the model contains a Dirac neutrino matrix $Y^{(0)}$ and that the light neutrino Majorana matrix $\mc M^{(0)}$ is diagonalized by the TBM matrix with an additional phase \cite{rrx2018asymmetric}. In this paper we construct $Y^{(-\frac{1}{3})}$ and $Y^{(-1)}$ through the introduction of gauge-singlet familons which spontaneously break the family symmetry. We postpone the discussion of the familon vacuum structure until all  familons contributing to the generation of the aforementioned mass matrices are known \cite{prrsx2019}.}

The paper is organized as follows. In Section \ref{sec:texture}, we revisit the key features of the asymmetric texture and seek a non-Abelian family symmetry that can naturally reproduce them. Section \ref{sec:T13} contains the relevant $\mc T_{13}$ group theory and a discussion on its merits for model building. \blue{In Section \ref{sec:effective}, we present an effective field theory model for constructing the asymmetric texture from a $\mc T_{13}$ family symmetry. The Higgs fields in our model are family-singlets, so that the matrix elements of the texture are generated from dimension-five and -six operators.} A theoretical outlook as to the origin of the $\mc T_{13}$ family symmetry follows in Section \ref{sec:theory}.

\section{A Family Symmetry for the Asymmetric Texture} \label{sec:texture}
\blue{The phenomenological asymmetric texture reproduces the deep ultraviolet structure of the Standard Model Yukawa matrices $Y^{(\frac{2}{3})}, Y^{(-\frac{1}{3})}$ and $Y^{(-1)}$.} \blue{Below we review its salient features, and show how it emerges as a minimal departure from symmetric Yukawa matrices in the context of $SU(5)$.}

\subsection{A Search for a Simple Texture}

Following the hints for ultraviolet simplicity outlined in the Introduction, an asymmetric texture for the down-quark and charged-lepton Yukawa matrices can be singled out under the following assumptions: 

\vskip .3cm
\noindent - Seesaw simplicity.  The two large leptonic mixing angles suggest that a good zeroth order approximation for $\mc U_{\rm PMNS}$ is TBM mixing.  We assume that 

\begin{equation} \label{tbmdelta}
\mc U_{Seesaw} = \mc U_{\rm TBM}(\delta) = \begin{pmatrix} \sqrt{\frac{2}{3}} & \frac{1}{\sqrt{3}} & 0 \\ 
-\frac{1}{\sqrt{6}} &  \frac{1}{\sqrt{3}} &  \frac{1}{\sqrt{2}} \\  \frac{e^{i \delta}}{\sqrt{6}} & - \frac{e^{i \delta}}{\sqrt{3}} &  \frac{e^{i \delta}}{\sqrt{2}} 
\end{pmatrix}.
\end{equation}
The addition of a phase in the third row serves to lower the corrections to the PMNS angles from the $\mc U^{(-1)}$ to their central PDG values. We assume that such mixing arises in the context of the seesaw mechanism, but do not further specify the dynamics of the Majorana sector; the origin of the phase $\delta$ and the implications of our chosen family symmetry on Majorana physics will be the focus of a future publication \cite{prrsx2019}. 

\vskip 0.3cm
\noindent - \blue{A diagonal up-quark Yukawa matrix }
 $Y^{({2\over 3})}=m_t\,{\rm Diag}(\lambda^8,\lambda^4,1)$, where we have expressed the mass ratios in terms of $\lambda$, the sine of the Cabibbo angle $\theta_{c}$. \blue{This feature of the asymmetric texture implies that the CKM matrix is generated by $Y^{(-\frac{1}{3})}$.} This is not a basis-dependent construction and needs to be explained by a symmetry.
 
\vskip .3cm
\noindent -   \blue{The $\gb{5}$ couplings of $Y^{(-{1\over 3})}_{}$ and $Y^{(-1)}_{}$ are related through transposition, as suggested by $SU(5)$. A $\overline{\g{45}}$ Higgs couples solely to the $(22)$ element of $Y^{(-\frac{1}{3})}$ and $Y^{(-1)}$, as in the Georgi-Jarlskog symmetric texture \cite{gj}.} The determinants of $Y^{(-{1\over 3})}_{}$ and  $Y^{(-1)}_{}$ are equal, i.e.,  the subdeterminant about their  $(22)$ matrix element vanishes.

\vskip 0.3cm
 
 \noindent With TBM mixing, purely symmetric or antisymmetric textures do not reproduce the data \cite{rrx2018asymmetric}. Some level of asymmetry is necessary in the Yukawa matrices to bring the reactor angle in agreement with its measured value. 
 
 For symmetric textures, 
\begin{equation} \label{eq:symmrot}
\mc U^{(-1)} = \mc U_{\rm CKM}(c \to -3c), 
\end{equation}
with $c$ the coupling of the $\overline{\g{45}}$ to the (22) position. \blue{With TBM neutrino mixing,} the correction to the leptonic mixing matrix yields a reactor angle
\begin{eqnarray} \nonumber
\lvert \sin \theta_{13} \rvert &=& \frac{1}{\sqrt{2}}\left| \mc U^{(-1)}_{21}+ \mc U^{(-1)}_{31} \right| 
\approx  \frac{\lambda}{3\sqrt{2}} = 0.051,
\end{eqnarray}
one third of its PDG value of 0.145. An asymmetric texture can alleviate this tension by relaxing Eq.~\eqref{eq:symmrot} to 
\begin{equation} \label{eq:asymmrot}
\mc U^{(-1)} = \mc V^{(-{1\over3})}(c \to -3c), 
\end{equation}
related now to the right-handed mixing of the down quarks.%

\blue{The phenomenological texture of \cite{rrx2018asymmetric} }, with a large asymmetry along the (13)-(31) axis, is given by

\be
Y^{(-{1\over3})} \sim 
\begin{pmatrix}
 b d \lambda ^4 & a \lambda ^3 & b \lambda ^3 \cr
 a \lambda ^3 & c \lambda ^2 & g \lambda ^2 \cr
 d \lambda  & g \lambda ^2 & 1 \end{pmatrix}
~~\mathrm{ and~~~}
Y^{(-1)} \sim 
\begin{pmatrix}
 b d \lambda ^4 & a \lambda ^3 & d \lambda \cr
 a \lambda ^3 & -3c \lambda ^2 & g \lambda ^2 \cr
 b \lambda^3  & g \lambda ^2 & 1 \end{pmatrix},
\ee
where $a$, $b$, $c$, $d$, and $g$ are $\mc O(1)$ prefactors, which serve as the input parameters to fit the experimental data. Written in terms of the Wolfenstein parameters $A$, $\rho$, and $\eta$, they are \cite{rrx2018asymmetric}
\begin{equation} \nonumber
a = c = \frac{1}{3}, ~~ g = A, ~~b = A \sqrt{\rho^2+\eta^2}, ~~d = \frac{2a}{g}=\frac{2}{3A}.
\end{equation}
This $\mc O(\lambda)$ asymmetry provides $\mc O(\lambda)$ elements to $\mc U^{(-1)}$, leading to 
\begin{equation} \nonumber
\sin \theta_{13} = \frac{\lambda}{3\sqrt{2}}\left(1+\frac{2}{A}\right) =0.184,
\end{equation} 
above its PDG value by $2.26^\circ$, with the solar and atmospheric angles also being slightly off their PDG values (by $\sim 3^\circ - 6^\circ$). \emph{All} leptonic mixing angles can be brought within $1^\circ$ of their central PDG values by the addition of a complex phase $\delta$ to the TBM mixing matrix, as in Eq.~\eqref{tbmdelta}.

\vskip 0.3cm

\noindent In summary, the key features of the asymmetric texture are:

\vskip .3cm
\noindent - a diagonal $Y^{({2\over 3})}_{}$;


\vskip .3cm
\noindent - an asymmetric $(31)$ and $(13)$ matrix element of $\m O(\lambda)$   \blue{in $Y^{(-\frac{1}{3})}$ and $Y^{(-1)}$, respectively, much larger than their transposed counterparts,} with symmetric off-diagonal elements elsewhere;

\vskip .3cm
\noindent - equality of the determinants of $Y^{(-{1\over3})}$ and $Y^{(-1)}$. This implies that the subdeterminant about the (22) entry of $Y^{(-{1\over3})}$ and $Y^{(-1)}$ must vanish.


\subsection{Asymmetric Group Theory}

\blue{The form of the Yukawa matrices in the asymmetric texture put strong constraints on the choice of a family symmetry group.}

\vskip 0.3cm
In $SU(5)$, the matter fields are described by three anti-quintets $F_i \sim \gb{5}$, and three decuplets $T_i \sim \g{10}$; \blue{we assume here that they transform as three-dimensional representations $\g{r}$ and $\g{s}$, respectively, of some family symmetry group, $G_f$.} 

The Yukawa matrices $Y^{(-\frac{1}{3})}$ and $Y^{(-1)}$ couple to $F \otimes T \equiv (\gb{5},\g{r}) \otimes (\g{10},\g{s})$. If these matrices are \emph{symmetric}, setting $\bf r=\bf s$ is natural, since group multiplication distinguishes symmetry from antisymmetry. In contrast, \emph{asymmetry} requires the identification of a specific off-diagonal matrix element, so that $\bf r$ and $\bf s$ must be \emph{different} representations:
\vskip 0.3cm
\red{{\bf{Requirement 1.}} $F$ and $T$ must be different triplets of $G_f$.}
\vskip 0.3cm
\blue{ The three smallest non-Abelian discrete subgroups of $SU(3)$ \cite{miller1961theory, *fairbairn1964finite, *ludl2011comments, *parattu2011tribimaximal, *grimus2014characterization, *Merle:2011vy, *Luhn:2011ip} with at least two \emph{distinct} three-dimensional representations are $\mc S_{4}$ of order $24$, $\Delta(27)$ of order $27$, and $\mc T_{13} = \mc Z_{13} \rtimes \mc Z_{3}$ of order $39$}. \blue{ $\mc S_{4}$ and $\Delta(27)$ have two real triplets, whereas $\mc T_{13}$ has two  complex triplets \cite{bovier1981finite, *bovier1981representations, *fairbairn1982some}.}
\vskip .3cm 
The diagonal charge-$2/3$ Yukawa matrix couples to  $T\otimes T \equiv ({\bf {10}},{\bf s})\otimes ({\bf  {10}},{\bf s})$, \blue{requiring:}
\vskip 0.3cm
\blue{{\bf{Requirement 2.}} The product  $\bf s\otimes\bf s$  distinguishes  diagonal from off-diagonal elements. }
\vskip 0.3cm
In $\mc S_4$, the product of a triplet with itself, i.e., 
\ben
\bf 3\otimes\bf 3=(\bf 1 \oplus \bf 2 \oplus \bf 3)_{s}\oplus \bf 3'_a,
\een
is such that the diagonal elements do not appear in a \emph{single} representation, irrespective of the choice of basis for Clebsch-Gordan coefficients \cite{ishimori2010non}. In $\Delta(27)$, the similar Kronecker product
\ben
\bf 3\otimes\bf 3=(\bf 3' \oplus \bf 3')_{s} \oplus \bf 3'_a,
\een
fails to put the diagonal elements in a \emph{distinct} triplet \cite{luhn2007flavor}. \blue{In both cases, singling out the diagonal elements requires some relations between coupling constants that are not protected by the group theory of either $\mc S_4$ or $\Delta(27)$}.  
\vskip 0.3cm
\noindent \blue{The group structure of $\mc T_{13}$ naturally satisfies the above requirements. } It yields a diagonal $Y^{(\frac{2}{3})}$ matrix, so that the CKM matrix is fully determined by the diagonalization of $Y^{(-\frac{1}{3})}$.


\section{$\mc T_{13}$ in a Nutshell} \label{sec:T13}
The two generators $a$ and $b$ of $\mc T_{13}=\mc{Z}_{13}\rtimes \mc Z_3$ have the presentation

\be
\langle a,b ~|~ a^{13}=b^3=I, bab^{-1}=a^3\rangle.
\ee
\red{The first two conditions establish $a$ and $b$ as generators of the $\mc Z_{13}$ and $\mc Z_{3}$ groups, while the third condition specifies how they nontrivially act under the semidirect product to construct $\mc T_{13}$.} 
Besides the trivial singlet $\bf 1$, $a$ and $b$ act on a complex one-dimensional irrep ${\bf 1'}$, two complex triplet irreps $\g{3}_1, \g{3}_2$, and their conjugates ${\bf \bar1'}$, $\gb{3}_1$ and $\gb{3}_2$. 

\red{In a simple choice of basis, the action of $a$ on triplets is to assign specific $\mc Z_{13}$ charges to the components, while $b$ cyclically permutes them. Thus,} the elements of each triplet can be labeled by mod $13$ arithmetic. Let $\rho^{13}=1$, and assign the charges as follows

\ben
{\bf 3_1}:~~~~(\rho,\,\rho^3,\,\rho^9),~~~~{\bf 3_2}:~~~~(\rho^2,\,\rho^6,\,\rho^5),
\een
with the mod $13$ conjugate charges in the conjugate representations. The Clebsch-Gordan coefficients are then determined by \red{the $\mc Z_{13}$ charges and the $\mc Z_3$ permutations.}

\vspace{0.3cm}
\noindent \red{For example,} setting  $\g{3}_2=\{\ket{1},\ket{2},\ket{3}\}$, we get under $\mc Z_3$,  

\ben
\ket{1}\rightarrow\ket{2}\rightarrow\ket{3}\rightarrow\ket{1},
\een
so that 
\ben
\ket{1}\ket{1}\rightarrow\ket{2}\ket{2}\rightarrow\ket{3}\ket{3}\rightarrow\ket{1}\ket{1}.
\een

\noindent Under $\mc Z_{13}$,  
\ben
\ket{1}\rightarrow\rho^2\ket{1},\quad \ket{2}\rightarrow\rho^6\ket{2},\quad \ket{3}\rightarrow\rho^5\ket{3},
\een
\ben
\ket{1}\ket{1}\rightarrow\rho^4\ket{1}\ket{1},\quad \ket{2}\ket{2}\rightarrow\rho^{12}\ket{2}\ket{2},\quad \ket{3}\ket{3}\rightarrow\rho^{10}\ket{3}\ket{3},
\een
which are exactly the charges of the $\bf \bar 3_1$ representation. This is reflected in the Kronecker product 
 
 \ben
\bf 3_2\otimes\bf 3_2=(\bf \bar 3_2 \oplus \bf \bar 3_1)_{s} \oplus (\bf \bar 3_{2})_a,
\een
with the diagonal elements in $\bf \bar 3_1$.  Similarly, in the Kronecker products

\begin{align*}
\bf 3_1\otimes\bf 3_1 = (\bf \bar 3_1 \oplus \bf 3_2)_{s} \oplus (\bf \bar 3_{1})_a , \quad \bf 3_1\otimes\bf 3_2 = \bf 3_2 \oplus \bf 3_1 \oplus  \bf \bar 3_{2},
\end{align*}
the diagonal elements reside in $\bf 3_2$ and $\bf 3_1$, respectively.

 The $\mc T_{13}$ group theory singles out the diagonal from off-diagonal elements, satisfying the first requirement: choosing the $SU(5)$ decuplet $T$ to transform as a triplet of $\mc T_{13}$, the up-quark matrix $Y^{({2\over 3})}_{}$ naturally appears diagonal, by which we mean that the relations between matrix elements are determined by the group structure.

To satisfy the second requirement, the antiquintets and decuplets must transform as distinct triplets of $\mc T_{13}$. Labeling their components as  $F=(F_1,\,F_2,\,F_3) \sim \g{3}_1$ and $T =(T_1,\,T_3,\,T_2)\sim \g{3}_2$, the tensor product yields (see Appendix~\ref{app:T13})

\begin{align} \label{eq:FT}
    \matc{
    F_1\\
    F_2\\
    F_3
    }_{\g{3}_1} \otimes
    \matc{
    T_1\\
    T_3\\
    T_2
    }_{\g{3}_2} &= 
    \matc{
    F_3 T_2 \\
    F_1 T_1 \\
    F_2 T_3
    }_{\g{3}_1} \oplus
    \matc{
    F_3 T_1\\
    F_1 T_3\\
    F_2 T_2
    }_{\gb{3}_2} \oplus
    \matc{
    F_3 T_3 \\
    F_1 T_2 \\
    F_2 T_1
    }_{\g{3}_2}.
\end{align}
Such an assignment of $\mc Z_{13}$ charges ensures that the three sets of symmetric off-diagonal matrix elements appear individually in the same representation, together with one diagonal element;  in this manner, $\mc T_{13}$ picks out individual matrix elements $F_i T_j$.

This assignment also sheds light on the construction of a diagonal $Y^{(\frac{2}{3})}$. For example, with $T$ transforming as a $\g{3}_2$, the dimension-five operator 
\be \rd{TTH_{\g{5}} \varphi^{(u)},}\ee
can generate the top-quark coupling with the simple vacuum alignment $\vev{\varphi^{(u)}} \sim (1,0,0)$\footnote{\rd{By $\vev{\varphi^{(u)}} \sim (1,0,0)$ we mean $\varphi^{(u)}_1\sim 1, \varphi^{(u)}_2=\varphi^{(u)}_3=0$.}}, \rd{where $\varphi^{(u)}$ is a familon field transforming as a $\g{3}_1$. The up- and charm-quark couplings may then be generated by higher dimensional operators, which may require additional familons as well as an extension of the $\mathcal{Z}_5$ shaping symmetry \cite{prrsx2019}. } As noted, $\mc T_{13}$ allows for such a diagonal construction, since the product of two similar triplets always picks out a unique familon representation for the diagonal couplings.

\blue{With the group and charge assignments determined, we now demonstrate how the key features of $Y^{(-\frac{1}{3})}$ and $Y^{(-1)}$ can be stitched together into a renormalizable theory.}
\section{Effective Theory Description} \label{sec:effective}
\blue{In our model, the Higgs fields $H_{\gb{5}} \sim \gb{5}$ and $H_{\overline{\g{45}}} \sim \overline{\g{45}}$ are $\mc T_{13}$ singlets, so that the Yukawa matrix elements are generated by effective operators of dimension five or higher. This requires the introduction of gauge-singlet familons $\varphi$ and $\varphi'$, which transform nontrivially under $\mc T_{13}$.}

\blue{ The dimension-five and -six effective operators  $FTH_{\gb{5}}\varphi$ and $FTH_{\gb{5}}\varphi \varphi'$, respectively, generate the $\gb{5}$ couplings. These operators can be constructed from renormalizable interactions by introducing a new complex messenger field $\Delta$, which yields the three vertices of Figure~\ref{fig:1}.}
\begin{figure}[ht!]
\centering
\subfloat[\label{ver_a}]{%
    	\begin{tikzpicture}[baseline=(a.base)]
    	\begin{feynman}
    	\vertex (a);
    	\vertex [right=of a] (c){\(\overline\Delta\)};
    	\vertex [above left=of a] (i1) {\(T\)};
    	\vertex [below left=of a] (i2) {\(H_{\gb{5}}\)};
    	\diagram* {
    		(i1) -- (a) -- (c),
    		(a) -- [scalar] (i2),
    	};
    	\end{feynman}
    	\end{tikzpicture}%
    }
\subfloat[\label{ver_b}]{%
        \begin{tikzpicture}[baseline=(a.base)]
        \begin{feynman}
        \vertex (a);
        \vertex [right=of a] (c){\({\Delta}\)};
        \vertex [above left=of a] (i1) {\(F\)};
        \vertex [below left=of a] (i2) {\(\varphi\)};
        \diagram* {
        	(i1) -- (a) -- (c),
        	(a) -- [scalar] (i2),
        };
        \end{feynman}
        \end{tikzpicture}%
    }
\subfloat[\label{ver_c}]{%
        \begin{tikzpicture}[baseline=(a.base)]
        \begin{feynman}
        \vertex (a);
        \vertex [right=of a] (c){\({\overline{\Delta}}\)};
        \vertex [above left=of a] (i1) {\(\Delta\)};
        \vertex [below left=of a] (i2) {\(\varphi'\)};
        \diagram* {
        	(i1) -- (a) -- (c),
        	(a) -- [scalar] (i2),
        };
        \end{feynman}
        \end{tikzpicture}%
    }
\caption{Vertices generating the effective Yukawa operators of the $\gb{5}$ couplings.}
    \label{fig:1}
\end{figure}

\vskip 0.2cm
 The vertex in Figure~\ref{ver_a} requires  $\overline\Delta$  to transform under $\mc T_{13}$ as a $\bf\overline 3_2$; the vertex in Figure~\ref{ver_b} implies that $\Delta$  transforms as a ${\bf 5}$ of $SU(5)$. By requiring $\Delta\sim({\bf 5},\bf 3_2)$, dimension-five interactions are  generated by $M_\Delta$, the invariant and presumably large  messenger mass. 
The vertex in Figure~\ref{ver_c} is possible because $\overline\Delta\Delta$ includes an $SU(5)$ singlet-$\mc T_{13}$ triplet term that couples to triplet familons. 

The relevant terms in the Lagrangian are of the form, 

\be
\label{lag1}
 y_0 T\overline{\Delta} H_{\gb{5}}+y F\Delta\varphi+M_{\Delta}\overline{\Delta}\Delta+y'\overline{\Delta}\Delta\varphi' \nonumber,
\ee
where $y, y'$ and $y_0$ are dimensionless coupling constants.
\vskip .2cm
\noindent The vertices in Figures \ref{ver_a} and \ref{ver_b} yield the following dimension-five interaction:

\begin{equation}\label{effectiveop}
	\centering
	\begin{tikzpicture}[baseline=(a.base)]
	\begin{feynman}[small]
	\vertex (a);
	\vertex [right=of a] (c);
	\vertex [above left=of a] (i1) {\(F\)};
	\vertex [below left=of a] (i2) {\(\varphi\)};
	\vertex [above right=of c] (f1) {\(T\)};
	\vertex [below right=of c] (f2) {\(H_{\gb{5}}\)};
	\diagram* {
		(i1) -- (a) --[insertion=0.5, edge label=\(\Delta\quad \overline{\Delta}\)] (c) -- (f1),
		(a) -- [scalar] (i2),
		(c) -- [scalar] (f2),
	};
	\end{feynman}
	\end{tikzpicture}
\end{equation}
\vskip .3cm
 \noindent whereas the vertex in Figure \ref{ver_c} is required to generate the following dimension-six interaction:

 \begin{equation}\label{eff2}
\begin{tikzpicture}[baseline=(a.base)]
\begin{feynman}[small]
\vertex (a);
\vertex [right=of a] (b);
\vertex [right=of b] (c);
\vertex [below=of b] (d) {$\varphi'$};
\vertex [above left=of a] (i1) {\(F\)};
\vertex [below left=of a] (i2) {\(\varphi\)};
\vertex [above right=of c] (f1)	 {\(T\)};
\vertex [below right=of c] (f2) {\(H_{\gb{5}}\)};
\diagram* {
	(i1) -- (a) --[insertion=0.5, edge label=\(\Delta\ \overline{\Delta}\)] (b) --[insertion=0.5, edge label=\(\Delta\ \overline{\Delta}\)] (c) -- (f1),
	(a) -- [scalar] (i2),
	(b) -- [scalar] (d),
	(c) -- [scalar] (f2),
};
\end{feynman}
\end{tikzpicture}
\end{equation}
\noindent Note that in Eq.~\eqref{lag1}, we have specifically chosen the operators so that $\varphi$ couples to $F$ and $H_{\gb{5}}$ couples to $T$.

 \sky{With the generic features of the effective operators explained, we now demonstrate how} the $\mc T_{13}$ Clebsch-Gordan coefficients enable us to separate out the asymmetric $(13)$ term and implement the zero subdeterminant with respect to the $(22)$ element.

\subsection{Generating the Asymmetric Term} \label{subsec:zero}
We obtain the asymmetric  $(31)$ matrix element, $Y^{(-{1\over 3})}_{31}$,  by dimension-five operators arising from the Lagrangian

\be\label{lag2}
\mc{L} ~\supset~ y_0 T\overline{\Delta} H_{\gb{5}}+y_2 F\Delta\varphi^{(2)}+ M_{\Delta}\overline{\Delta}\Delta,
\ee
where the familon  $\varphi^{(2)}$ transforms as a $\g{3}_2$ of $\mc T_{13}$, with vacuum alignment

\begin{align*}
\vev{\varphi^{(2)}} &~ \text{ along }~ (0, 1, 0).
\end{align*}
It yields the $F_1T_3$ entry after integrating out the heavy messenger fields $\Delta$ and $\overline{\Delta}$, i.e.,

\begin{align}
\frac{1}{M_\Delta}FTH_{\gb{5}}\varphi^{(2)}~&\rightarrow~~~ \dfrac{y_0y_2\vev{H_{\gb{5}}}\vev{\varphi^{(2)}_2}}{M_\Delta}F_1T_3\label{e13}\\
\nonumber\\
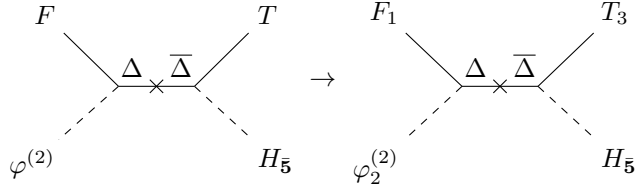
\begin{tikzpicture}[baseline=(a.base)]
\begin{feynman}[small]
\vertex (a);
\vertex [right=of a] (c);
\vertex [above left=of a] (i1) {\(F\)};
\vertex [below left=of a] (i2) {\(\varphi^{(2)}\)};
\vertex [above right=of c] (f1) {\(T\)};
\vertex [below right=of c] (f2) {\(H_{\gb{5}}\)};
\diagram* {
	(i1) -- (a) --[insertion=0.5, edge label=\(\Delta\quad \overline{\Delta}\)] (c) -- (f1),
	(a) -- [scalar] (i2),
	(c) -- [scalar] (f2),
};
\end{feynman}
\end{tikzpicture}&\rightarrow
\begin{tikzpicture}[baseline=(a.base)]
\begin{feynman}[small]
\vertex (a);
\vertex [right=of a] (c);
\vertex [above left=of a] (i1) {\(F_1\)};
\vertex [below left=of a] (i2) {\(\varphi^{(2)}_2\)};
\vertex [above right=of c] (f1) {\(T_3\)};
\vertex [below right=of c] (f2) {\(H_{\gb{5}}\)};
\diagram* {
	(i1) -- (a) --[insertion=0.5, edge label=\(\Delta\quad \overline{\Delta}\)] (c) -- (f1),
	(a) -- [scalar] (i2),
	(c) -- [scalar] (f2),
};
\end{feynman}
\end{tikzpicture}\nonumber
\end{align}
Here $\varphi^{(a)}_i$ corresponds to the $i^{\mathrm{th}}$ component of the triplet $\varphi^{(a)}$.
\vskip .3cm
The diagonal $(33)$ element $Y^{(-{1\over 3})}_{33}$ can similarly be generated by adding the term $y_1 F\Delta \varphi^{(1)}$ to the Lagrangian of Eq.~\eqref{lag2}. It requires a familon $\varphi^{(1)} \sim \gb{3}_2$ with vacuum alignment 
\begin{align*}
\vev{\varphi^{(1)}}~~~\text{along}~~~(1,0,0).
\end{align*}
Integrating out $\Delta$ and $\overline{\Delta}$ gives rise to the effective operator $FTH_{\gb{5}}\varphi^{(1)}$, yielding the desired term

\begin{align}
\frac{1}{M_\Delta}FTH_{\gb{5}}\varphi^{(1)}~&\rightarrow~~~ \dfrac{y_0y_1\vev{H_{\gb{5}}}\vev{\varphi^{(1)}_1}}{M_\Delta} F_3T_3\label{e33}\\
&~\nonumber\\
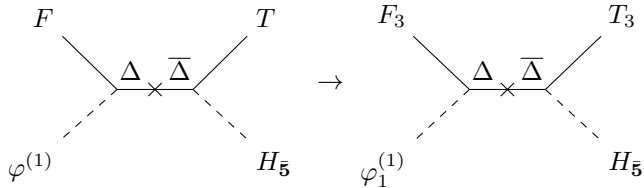
\begin{tikzpicture}[baseline=(a.base)]
\begin{feynman}[small]
\vertex (a);
\vertex [right=of a] (c);
\vertex [above left=of a] (i1) {\(F\)};
\vertex [below left=of a] (i2) {\(\varphi^{(1)}\)};
\vertex [above right=of c] (f1) {\(T\)};
\vertex [below right=of c] (f2) {\(H_{\gb{5}}\)};
\diagram* {
	(i1) -- (a) --[insertion=0.5, edge label=\(\Delta\quad \overline{\Delta}\)] (c) -- (f1),
	(a) -- [scalar] (i2),
	(c) -- [scalar] (f2),
};
\end{feynman}
\end{tikzpicture}
~&\rightarrow
\begin{tikzpicture}[baseline=(a.base)]
\begin{feynman}[small]
\vertex (a);
\vertex [right=of a] (c);
\vertex [above left=of a] (i1) {\(F_3\)};
\vertex [below left=of a] (i2) {\(\varphi^{(1)}_1\)};
\vertex [above right=of c] (f1) {\(T_3\)};
\vertex [below right=of c] (f2) {\(H_{\gb{5}}\)};
\diagram* {
	(i1) -- (a) --[insertion=0.5, edge label=\(\Delta\quad \overline{\Delta}\)] (c) -- (f1),
	(a) -- [scalar] (i2),
	(c) -- [scalar] (f2),
};
\end{feynman}
\end{tikzpicture}\nonumber
\end{align}

\sky{With the $(33)$ term and the asymmetric $(31)$ term constructed by dimension-five effective operators, we next show how to generate the vanishing of the subdeterminant from $\mc T_{13}$ group structure. }

\subsection{Generating the Zero Subdeterminant}
The asymmetric texture requires the vanishing of the subdeterminant about $Y^{(-{1\over 3})}_{22}$ and $Y^{(-1)}_{22}$. It implies that the $(1\text{-}3)$ submatrix takes the form

\be\label{2*2}
	\begin{pmatrix}
	\gamma\alpha & \gamma\beta\\
	\alpha & \beta
	\end{pmatrix}. \nonumber
	\ee
The \emph{first} row matrix elements, of $\mc O(\lambda^4)$  and $\mc O(\lambda^3)$ respectively, are much smaller than those of the \emph{second} row  ($\mc O(\lambda)$  and $\mc O(1)$). This is a unique feature of the asymmetric texture, in contrast to the symmetric Georgi-Jarlskog texture \cite{gj}. It suggests that the upper-row elements of the $(1\text{-}3)$ submatrix are generated  by six (or higher) dimensional effective operators.
\vskip 0.2cm

To generate $Y^{(-{1\over 3})}_{13}$ and $Y^{(-{1\over 3})}_{11}$, we add a new interaction $y_3\overline{\Delta}\Delta\varphi^{(3)}$ to Eq.~\eqref{lag2}, yielding the following dimension-six operators:
\begin{align}
\frac{1}{M_\Delta^2}FTH_{\gb{5}}\varphi^{(1)}\varphi^{(3)} ~&\rightarrow~~~ \dfrac{y_0y_1y_3\vev{H_{\gb{5}}}\vev{\varphi^{(1)}_1} \vev{\varphi^{(3)}_k}}{M_\Delta^2}F_3T_1\label{e31}\\
\nonumber\\
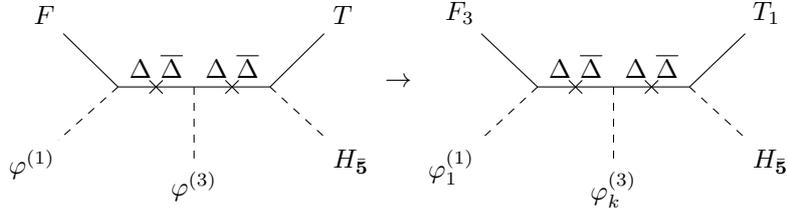
\begin{tikzpicture}[baseline=(a.base)]
\begin{feynman}[small]
\vertex (a);
\vertex [right=of a] (b);
\vertex [right=of b] (c);
\vertex [below=of b] (d) {$\varphi^{(3)}$};
\vertex [above left=of a] (i1) {\(F\)};
\vertex [below left=of a] (i2) {\(\varphi^{(1)}\)};
\vertex [above right=of c] (f1)	 {\(T\)};
\vertex [below right=of c] (f2) {\(H_{\gb{5}}\)};
\diagram* {
	(i1) -- (a) --[insertion=0.5, edge label=\(\Delta\ \overline{\Delta}\)] (b) --[insertion=0.5, edge label=\(\Delta\ \overline{\Delta}\)] (c) -- (f1),
	(a) -- [scalar] (i2),
	(b) -- [scalar] (d),
	(c) -- [scalar] (f2),
};
\end{feynman}
\end{tikzpicture} &\rightarrow
\begin{tikzpicture}[baseline=(a.base)]
\begin{feynman}[small]
\vertex (a);
\vertex [right=of a] (b);
\vertex [right=of b] (c);
\vertex [below=of b] (d) {$\varphi^{(3)}_k$};
\vertex [above left=of a] (i1) {\(F_3\)};
\vertex [below left=of a] (i2) {\(\varphi^{(1)}_1\)};
\vertex [above right=of c] (f1) {\(T_1\)};
\vertex [below right=of c] (f2) {\(H_{\gb{5}}\)};
\diagram* {
	(i1) -- (a) --[insertion=0.5, edge label=\(\Delta\ \overline{\Delta}\)] (b) --[insertion=0.5, edge label=\(\Delta\ \overline{\Delta}\)] (c) -- (f1),
	(a) -- [scalar] (i2),
	(b) -- [scalar] (d),
	(c) -- [scalar] (f2),
};
\end{feynman}
\end{tikzpicture}\nonumber
\end{align}
and
\begin{align}
\frac{1}{M_\Delta^2}FTH_{\gb{5}}\varphi^{(2)}\varphi^{(3)} ~&\rightarrow~~~ \dfrac{y_0y_2y_3 \vev{H_{\gb{5}}}\vev{\varphi^{(2)}_2} \vev{\varphi^{(3)}_k}}{M_\Delta^2}F_1T_1\label{e11}\\
\nonumber \\
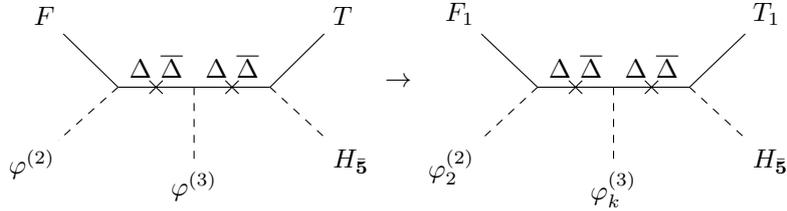
\begin{tikzpicture}[baseline=(a.base)]
\begin{feynman}[small]
\vertex (a);
\vertex [right=of a] (b);
\vertex [right=of b] (c);
\vertex [below=of b] (d) {$\varphi^{(3)}$};
\vertex [above left=of a] (i1) {\(F\)};
\vertex [below left=of a] (i2) {\(\varphi^{(2)}\)};
\vertex [above right=of c] (f1) {\(T\)};
\vertex [below right=of c] (f2) {\(H_{\gb{5}}\)};
\diagram* {
	(i1) -- (a) --[insertion=0.5, edge label=\(\Delta\ \overline{\Delta}\)] (b) --[insertion=0.5, edge label=\(\Delta\ \overline{\Delta}\)] (c) -- (f1),
	(a) -- [scalar] (i2),
	(b) -- [scalar] (d),
	(c) -- [scalar] (f2),
};
\end{feynman}
\end{tikzpicture} &\rightarrow
\begin{tikzpicture}[baseline=(a.base)]
\begin{feynman}[small]
\vertex (a);
\vertex [right=of a] (b);
\vertex [right=of b] (c);
\vertex [below=of b] (d) {$\varphi^{(3)}_k$};
\vertex [above left=of a] (i1) {\(F_1\)};
\vertex [below left=of a] (i2) {\(\varphi^{(2)}_2\)};
\vertex [above right=of c] (f1) {\(T_1\)};
\vertex [below right=of c] (f2) {\(H_{\gb{5}}\)};
\diagram* {
	(i1) -- (a) --[insertion=0.5, edge label=\(\Delta\ \overline{\Delta}\)] (b) --[insertion=0.5, edge label=\(\Delta\ \overline{\Delta}\)] (c) -- (f1),
	(a) -- [scalar] (i2),
	(b) -- [scalar] (d),
	(c) -- [scalar] (f2),
};
\end{feynman}
\end{tikzpicture}\nonumber
\end{align}
The required $\mc {Z}_{13}$ charge of $\varphi^{(3)}_k$ is $\rho^4$, which implies that $\varphi^{(3)}$ must transform as a $\gb{3}_1$, with vacuum alignment
\begin{align*}
\vev{\varphi^{(3)}} ~~~\text{along}~~~ (0,0,1). 
\end{align*}

The matrix elements of the submatrix are then given by

\bean
Y^{(-{1\over 3})}_{11}=\frac{y_0y_2y_3\langle H_{\gb{5}}\rangle\langle\varphi^{(2)}_2\rangle\langle\varphi^{(3)}_3\rangle}{M_\Delta^2},~~~
Y^{(-{1\over 3})}_{13}&=&\frac{y_0y_1y_3\langle H_{\gb{5}}\rangle\langle\varphi^{(1)}_1\rangle\langle\varphi^{(3)}_3\rangle}{M_\Delta^2},\\
Y^{(-{1\over 3})}_{31}=\frac{y_0y_2\langle H_{\gb{5}}\rangle\langle\varphi^{(2)}_2\rangle}{M_\Delta},~~~~~~~~~~~~~
 Y^{(-{1\over 3})}_{33}&=&\frac{y_0y_1\langle H_{\gb{5}}\rangle\langle\varphi^{(1)}_1\rangle}{M_\Delta}.
\eean
They naturally generate the desired  `zero subdeterminant' condition, \emph{independently} of the coupling constants $y_i$, i.e., 
\begin{align}
{Y^{(-{1\over 3})}_{11} Y^{(-{1\over 3})}_{33} = Y^{(-{1\over 3})}_{13} Y^{(-{1\over 3})}_{31} }.
\end{align}
Its implementation is possible courtesy of the $\mc T_{13}$ Clebsch-Gordan coefficients and the choice of vertices in Figure~\ref{fig:1}. If instead we had chosen $H_{\gb{5}}$ to couple to $F$ and $\varphi$ to couple to $T$, as in $F\overline{\Delta}H_{\gb{5}}$, $T\Delta \varphi^{(1)}$ and $T\Delta \varphi^{(2)}$, the zero subdeterminant condition could not have been implemented. 

 \vskip .5cm
The remaining symmetric off-diagonal elements can be generated by adding two  familons, $\varphi^{(4)} \sim \gb{3}_2$ and $\varphi^{(5)} \sim \gb{3}_1$, contributing two more terms, $y_4 F \Delta \varphi^{(4)}$ and $y_5 F \Delta \varphi^{(5)}$, to the Lagrangian of Eq.~\eqref{lag2}. The required vacuum alignment for the familons are $(0,1,1)$ and $(1,0,1)$, respectively.  
\vskip 0.5 cm

\sky{The last required feature of the texture is the generation of the $(22)$ element by the $\overline{\g{45}}$ coupling, which we turn to next.}

\subsection{The $\overline{\g{45}}$ Coupling} \label{subsec:45}
The $(22)$ term is solely generated by the coupling to a Higgs $H_{\overline{\g{45}}}$ transforming as a $\overline{\g{45}}$ of $SU(5)$. The invariant in terms of $SU(5)$ indices $a,b,c$ is $F_aT^{bc}{H_{\overline{\g{45}}}}_{bc}^a$. 
For simplicity, we consider this Higgs to be a singlet of $\mc T_{13}$. A familon ${\varphi^{(6)}}$ generates the $(22)$ term with a dimension-five effective operator of the form
\begin{align*}
 \frac{1}{\Lambda} F T H_{\overline{\g{45}}} {\varphi^{(6)}}.
\end{align*}
From Eq.~\eqref{eq:FT}, we require that ${\varphi^{(6)}}$ transforms as a $\g{3}_2$, aligned along the $(0,0,1)$ direction in the vacuum.

\vspace{0.2cm}
 At tree level, this effective operator can be constructed by introducing a \emph{new} complex ``messenger'' field $\Sigma$ with heavy mass $M_\Sigma$. Consider the scenario where the Higgs couples to $F$ and the familon couples to $T$, as in Figure~\ref{fig:ver45}.

\begin{figure}[ht!]
    \centering
    \subfloat[]{
        \begin{tikzpicture}[baseline=(a.base)]
        \begin{feynman}
        \vertex (a);
        \vertex [right=of a] (c){\(\overline{\Sigma}\)};
        \vertex [above left=of a] (i1) {\(F\)};
        \vertex [below left=of a] (i2) {\(H_{\overline{\g{45}}}\)};
        \diagram* {
        	(i1) -- (a) -- (c),
        	(a) -- [scalar] (i2),
        };
        \end{feynman}
        \end{tikzpicture} \label{ver_a45}
    }
    \subfloat[]{
        \begin{tikzpicture}[baseline=(a.base)]
        \begin{feynman}
        \vertex (a);
        \vertex [right=of a] (c){\(\Sigma\)};
        \vertex [above left=of a] (i1) {\(T\)};
        \vertex [below left=of a] (i2) {\({\varphi^{(6)}}\)};
        \diagram* {
        	(i1) -- (a) -- (c),
        	(a) -- [scalar] (i2),
        };
        \end{feynman}
        \end{tikzpicture} \label{ver_b45}
    }
    \caption{Vertices generating the effective Yukawa operator of the $\overline{\g{45}}$ coupling.}
    \label{fig:ver45}
\end{figure}
\noindent From Figure~\ref{ver_a45}, $\Sigma \sim \g{3}_1$ of $\mc T_{13}$, and from Figure~\ref{ver_b45}, $\Sigma \sim \overline{\g{10}}$ of $SU(5)$. 

A Lagrangian of the form

\begin{align} \label{lag45}
	\mathcal{L}_{\overline{\g{45}}} &= y_6 F\overline{\Sigma} H_{\overline{\g{45}}}+ y_7 T\Sigma {\varphi^{(6)}}+M_\Sigma \overline{\Sigma} \Sigma
\end{align} 
yields the requisite operator:

\begin{align}
\frac{1}{M_\Sigma} FTH_{\overline{\g{45}}}{\varphi^{(6)}} ~~~&\rightarrow~~~ \frac{y_7 y_8 \langle H_{\overline{\g{45}}}\rangle\langle\varphi^{(6)}_3\rangle}{M_\Sigma}F_2T_2 .\\
\begin{tikzpicture}[baseline=(a.base)]
\begin{feynman}[small]
\vertex (a);
\vertex [right=of a] (c);
\vertex [above left=of a] (i1) {\(F\)};
\vertex [below left=of a] (i2) {\(H_{\overline{\g{45}}}\)};
\vertex [above right=of c] (f1) {\(T\)};
\vertex [below right=of c] (f2) {\({\varphi^{(6)}}\)};
\diagram* {
	(i1) -- (a) --[insertion=0.5, edge label=\(\overline{\Sigma}\quad \Sigma\)] (c) -- (f1),
	(a) -- [scalar] (i2),
	(c) -- [scalar] (f2),
};
\end{feynman}
\end{tikzpicture} ~~~ &\rightarrow ~~~
\begin{tikzpicture}[baseline=(a.base)]
\begin{feynman}[small]
\vertex (a);
\vertex [right=of a] (c);
\vertex [above left=of a] (i1) {\(F_2\)};
\vertex [below left=of a] (i2) {\(H_{\overline{\g{45}}}\)};
\vertex [above right=of c] (f1) {\(T_2\)};
\vertex [below right=of c] (f2) {\(\varphi^{(6)}_3\)};
\diagram* {
	(i1) -- (a) --[insertion=0.5, edge label=\(\overline{\Sigma}\quad \Sigma\)] (c) -- (f1),
	(a) -- [scalar] (i2),
	(c) -- [scalar] (f2),
};
\end{feynman}
\end{tikzpicture}\nonumber
\end{align}

\sky{This completes the description of the Yukawa couplings.}

\subsection{The Familon Vacuum}
$\mc T_{13}$ Clebsch-Gordan coefficients determine the matrix elements of the asymmetric texture, with the aid of six familons. The representations and vacuum alignments of these familons are

\bean
\varphi^{(1)} \sim \gb{3}_2~:~\vev{H_{\gb{5}}}\langle\varphi^{(1)}\rangle &\sim& M_\Delta(1,0,0),\\
\varphi^{(2)} \sim  \g{3}_2~:~\vev{H_{\gb{5}}}\langle\varphi^{(2)}\rangle &\sim& d\lambda~ M_\Delta(0,1,0),\\
\varphi^{(3)} \sim \gb{3}_1~:~\vev{H_{\gb{5}}}\langle\varphi^{(3)}\rangle &\sim& b\lambda^3 M_\Delta(0,0,1),\\
\varphi^{(4)} \sim \gb{3}_2~:~\vev{H_{\gb{5}}}\langle\varphi^{(4)}\rangle &\sim& a\lambda^3 M_\Delta(0,1,1),\\
\varphi^{(5)} \sim \gb{3}_1~:~\vev{H_{\gb{5}}}\langle\varphi^{(5)}\rangle &\sim& g\lambda^2 M_\Delta(1,0,1),\\
\varphi^{(6)} \sim  \g{3}_2~:\vev{H_{\overline{\g{45}}}}\langle\varphi^{(6)}\rangle &\sim& c\lambda^2 M_\Sigma(0,0,1).
\eean
\blue{These vacuum directions have a geometric feature, in the sense that they resemble the sides and face-diagonals of a cube. $\mc T_{13}$ assigns to each familon component a unique $\mc Z_{13}$ charge, allowing them to pick out specific matrix elements.}


\subsection{Extending $SU(5) \times \mc T_{13}$ with an Abelian $\mc Z_{5}$ Symmetry} \label{subsec:danger}
As we have seen, this familon structure enables an elegant $SU(5) \times \mc T_{13}$ model of the asymmetric Yukawa texture given in \cite{rrx2018asymmetric}. However, the fact that some of the familons belong to the same representation means that unwanted couplings can be generated. This results in a need to extend $SU(5) \times \mc T_{13}$ by an additional symmetry to protect against such couplings.

\blu{More precisely, the necessity to extend the $SU(5) \times \mc T_{13}$ symmetry stems from the need to: (a) separate the $\gb{5}$ and $\overline{\g{45}}$ couplings, and  (b) prevent additional operators to which the familons could couple inadvertently. }

\vskip 0.2cm
 --(a) \sky{The $\gb{5}$ and $\overline{\g{45}}$ couplings do not mix in the asymmetric texture. One could implement the $\overline{\g{45}}$ coupling with the same messenger $\Delta$ used for the $\gb{5}$ couplings, where $H_{\overline{\g{45}}}$ couples to $T$ and ${\varphi^{(6)}}$ couples to $F$. However, this contributes an unwanted $H_{\gb{5}}$ coupling to $Y^{(-\frac{1}{3})}_{22}$. It can be avoided by introducing a new symmetry under which $H_{\gb{5}}$ and $H_{\overline{\g{45}}}$ transform differently, thus requiring a new messenger field $\Sigma$. }

\vspace{0.2cm}
 --(b) \sky{The second reason for extending the symmetry arises from the familons being complex fields, with some of them having same transformation properties under $SU(5) \times \mc T_{13}$. For example, both $\varphi^{(3)}$ and $\varphi^{(5)}$ transform as a $\gb{3}_1$ of $\mc T_{13}$, allowing
the term $y'_5 ~\overline{\Delta} \Delta \varphi^{(5)}$.} Together with the terms $y_1 F\Delta \varphi_1$ and $y_0 T\overline{\Delta} H_{\gb{5}}$, it yields the dimension-six operator,

\begin{align*}
\frac{1}{M_\Delta^2}FTH_{\gb{5}}\varphi^{(1)}\varphi^{(5)} ~&\rightarrow~~~ \dfrac{y_0 y_1 y'_5\vev{H_{\gb{5}}}\vev{\varphi^{(1)}_1} \vev{\varphi^{(5)}_1}}{M_\Delta^2}F_1 T_1 \nonumber \\
&+~~~~  \dfrac{y_0 y_1 y'_5\vev{H_{\gb{5}}}\vev{\varphi^{(1)}_1} \vev{\varphi^{(5)}_3}}{M_\Delta^2}F_3 T_1,
\end{align*}

\noindent contributing $g\lambda^2$ to $Y^{(-\frac{1}{3})}_{11}$ and $Y^{(-\frac{1}{3})}_{13}$, larger than the required leading terms of $\mc O(\lambda^4)$ and $\mc O(\lambda^3)$. Consider another example, with $\varphi^{(2)} \sim \g{3}_2$, $\varphi^{(2)*} \sim \gb{3}_2$. The allowed term $F\Delta\varphi^{(2)*}$ would contribute an $\mc O(\lambda)$ term to  $Y^{(-\frac{1}{3})}_{21}$, larger than the desired $\mc O(\lambda^3)$ term.

\vskip 0.2cm
All such problems can be alleviated by introducing a new symmetry and carefully choosing the charges of the fields. The \emph{smallest} group which works is $\mc Z_5$, as we show in Appendix~\ref{app:shape}. 

\blu{The full symmetry of the down-quark and charged-lepton sectors is therefore $SU(5) \times \mc T_{13} \times \mc Z_{5}$.} 

The transformation properties of the fields are listed in Table~\ref{table:summary}.

\begin{table}[ht]\centering
\renewcommand\arraystretch{2}
\begin{tabularx}{\textwidth}{@{}l | Y Y Y Y Y Y | Y Y Y Y Y Y @{}}
\toprule
     & $F$ & $T$ & $H_{\gb{5}}$ & $H_{\overline{\g{45}}}$ & $\Delta$  & $\Sigma$ & $\varphi^{(1)}$ & $\varphi^{(2)}$ & $\varphi^{(3)}$ & $\varphi^{(4)}$ & $\varphi^{(5)}$&${\varphi^{(6)}}$\\ 
\hline
$SU(5)$    & $\overline{ \g{5}}$ & $\g{10}$ & $\overline{ \g{5}}$&$\overline{\g{45}}$ & $\g{5}$ & $\overline{ \g{10}}$ & $\g{1}$ & $\g{1}$ & $\g{1}$ & $\g{1}$ & $\g{1}$ & $\g{1}$ \\ 

$\mc T_{13}$ &  $\g{3}_1$ & $\g{3}_2$ & $\g{1}$ & $\g{1}$ & $\g{3}_2$  & $\g{3}_1$ & $\gb{3}_2$ & $\g{3}_2$ & $\gb{3}_1$  & $\gb{3}_2$ & $\gb{3}_1$ & $\g{3}_2$\\ 

$\mathcal{Z}_5$    & $\g{1}$ & $\g{1}$ & $\g{\eta^4}$& $\g{\eta^3}$ & $\g{\eta^4}$  & $\g{\eta^3}$ & $\g{\eta}$   & $\g{\eta}$ & $\g{1}$  & $\g{\eta}$ & $\g{\eta}$ & $\g{\eta^2}$ \\ 
		
\botrule
\end{tabularx} 
\caption{Charge assignments of matter, Higgs, messenger and familon fields ($\eta=e^{\frac{2\pi i}{5}}$).}
\label{table:summary}
\end{table}

Note that the familons $\varphi^{(1)}$ and $\varphi^{(4)}$ still have the same transformation properties, although the vev of $\varphi^{(4)}$ is suppressed by a factor of $\mc O(\lambda^3)$. In principle, they should couple to the same fields, and they do so in our model; they should also mix. However, as their vacuum alignments are orthogonal, \rd{this has no effect on the asymmetric texture.} 

The full Yukawa Lagrangian generating the matrix elements of the phenomenological asymmetric texture is given by
\begin{align}
\mathcal{L}_Y &= \mc L_{\gb{5}} + \mc L_{\overline{\g{45}}} \nonumber \\
    &= y_0 T\overline{\Delta} H_{\gb{5}}+y_1 F\Delta\varphi^{(1)} +y_2 F\Delta\varphi^{(2)}+M_{\Delta}\overline{\Delta}\Delta+y_3\overline{\Delta}\Delta\varphi^{(3)}+y_4 F\Delta\varphi^{(4)} \nonumber \\
 &+y_5 F\Delta\varphi^{(5)}
+y_6 F\overline{\Sigma} H_{\overline{\g{45}}}+ y_7 T\Sigma {\varphi^{(6)}}+M_\Sigma \overline{\Sigma} \Sigma . \label{full_lagrang}
\end{align}



\section{Theoretical Outlook}\label{sec:theory}

In the $\mc T_{13}$ model, asymmetry arises naturally only when $F$ and $T$ transform as different family triplets. This might seem counterintuitive in a theory that relies on gauge unification. Yet, it may not be so odd at the level of $E_6$.

\vskip 0.2cm
The  $E_6$ fundamental representation  decomposes as $\bf 27 = \bf 16 \oplus 10 \oplus 1 = [\bar 5 \oplus   10 \oplus 1] \oplus [5 \oplus \bar 5] \oplus 1$ under $SO(10)$ and $SU(5)$, respectively.  The $SU(5)$ $\bf 5$ in the $SO(10)$ decuplet could acquire a vectorlike mass by coupling with  the $\bf \bar 5$ in the $SO(10)$ $\bf 16$. The chiral content would then be  $SU(5)$ $\bf 10$s and  $\bf\bar 5$s coming from different representations:
$$
SO(10)\times \mc T_{13}:~~~(\bf 16, 3_2) \oplus (10,3_1).
$$

\vskip 0.3cm
$\mc T_{13}$ and $\mc T_7$  \cite{t7luhn, *muterm} are well known to physicists as discrete subgroups of the continuous group $SU(3)$ since they have three-dimensional complex representations. \blu{They have also been discussed in connection to the global symmetries of two-dimensional spin lattice models, where each lattice point has a $\mc Z_{7}\times \mc Z_{3}$ and $\mc Z_{13}\times \mc Z_{3}$ symmetry, respectively, and the direct product becomes a semidirect product for special values of the interaction strength between nearest neighbors \cite{marcu1981global1, *marcu1981global2, *rittenberg1982global}.}

They are also subgroups of the continuous group $G_2$ \cite{cohen1983finite, *cohen2014spectral, *king1999finite} through two different embeddings. In one, they are subgroups of $SU(3)$, which is a subgroup of $G_2$. In the other, the embedding goes through the seven-dimensional representation of $G_2$, bypassing $SU(3)$. For $\mc T_7$, this sequence is 
$$\mc T_7=\mc Z_7\rtimes \mc Z_3~\subset~ \mc P\mc S\mc L(2,7)~\subset~G_2,$$
where the seven-dimensional representation  of $\mc P\mc S\mc L(2,7)$ is equal to that of continuous $G_2$. 
The same septet embedding is also present for $\mc T_{13}$:
$$\mc T_{13}=\mc Z_{13}\rtimes \mc Z_3~\subset~\mc Z_{13}\rtimes \mc Z_6~\subset \mc P\mc S\mc L(2,13)~\subset~G_2.$$ 
This case is more complicated since $ \mc P\mc S\mc L(2,13)$ has two distinct septet representations. In either case, these ``anomalous'' embeddings single out a seven-dimensional manifold. Applied to compactification, it could point to eleven-dimensional physics.

\section{Conclusions}

In this paper we presented a family symmetry model based on the group $\mc T_{13}$  to derive the asymmetric texture proposed in an earlier work.  The key features of the asymmetry  are well explained by $\mc T_{13}$. \red{With a simple choice of basis inspired by mod 13 arithmetic,} its Clebsch-Gordan coefficients naturally single out diagonal matrices; this feature is crucial for the charge-$2/3$ Yukawa matrix. The $SU(5)$ fermion fields $F$ and $T$ transform as distinct $\mc T_{13}$ triplets, distinguishing each matrix element. By relabeling $T$ as $(T_1,T_3,T_2) \sim{\bf 3_2}$, and keeping $F$ as $(F_1,F_2,F_3) \sim {\bf 3_1}$, the symmetric terms, $F_i T_j$ and $F_j T_i$, appear in the same triplets. Six Yukawa couplings of the down quarks and charged leptons are generated by dimension-five effective operators obtained by integrating out a complex massive messenger field $\Delta$. Two couplings are described  by dimension-six effective operators.

\vskip 0.2cm 
With  $\mc T_{13}$, the equality of the determinants of the down-quark and charged-lepton matrices required by GUT-scale mass ratios is satisfied without fine-tuning. The Georgi-Jarlskog ${\overline{\g{45}}}$ coupling in the $(22)$ position is given by another dimension-five operator, generated by integrating out a different complex messenger field $\Sigma$. \sky{An abelian symmetry, $\m Z_5$, is needed to distinguish the messengers of $\gb{5}$ and $\overline{\g{45}}$ couplings and label the familons to restrict unwanted terms in the tree-level Lagrangian.} 


\vskip 0.2cm 
\sky{The model presented in this paper addresses only the down-quark and charged-lepton Yukawa matrices of the Standard Model. It serves as a small step towards a more complete model, requiring additional symmetries and familon fields, that addresses all mass matrices and familon dynamics. When applied to the neutrino sector with a complex TBM mixing, it reproduces the observable mixing angles and predicts leptonic CP violation. However, it }does not resolve the  ordering of the light neutrino masses nor does it specify the underlying dynamics of the neutrino sector. The origin of the phase in the TBM matrix is still unknown. Perhaps it can be generated from a generalized CP symmetry \cite{everett2017lepton, *ding2013generalised, *li2016a4, *nishi2013generalized, *sinha2019c, *ding2014lepton, *hagedorn2015lepton, *ding2016generalized, *li2015lepton, *di2015lepton, *ballett2015mixing, *turner2015predictions, *ding2013spontaneous, *feruglio2013lepton,  *feruglio2014realistic, *li2014generalised, *li2015deviation, *lu2017alternative, *king2014lepton, *ding2014generalized, *ding2014generalised2, *chen2015neutrino, *chen2018neutrino, *lu2018quark, *joshipura2018pseudo, *rong2017lepton, *Everett:2015oka, *Lu:2018oxc, *girardi2014generalised, *penedo2017neutrino, *Li:2017abz, *Li:2017zmk, *Lu:2016jit, *Li:2016nap, *Yao:2016zev, *Chen:2016ica, *Chen:2016ptr, *Li:2016ppt, *Ding:2013nsa, *chen2019cp, *chen2019cp2, *barreiros2019combining, *sinha2019phenomenological}. 

The asymmetric texture together with the complex TBM mixing can also predict Majorana invariants, from which one can calculate the Majorana phases and express the effective Majorana mass parameter $m_{\beta\beta}$ of neutrinoless double beta decay in terms of the lightest neutrino mass. Extending the $\mc T_{13}$ model to the neutrino sector, one can thus predict the light neutrino masses, and  $m_{\beta \beta}$, with an additional constraint coming from $\mc T_{13}$ invariants. Also, the familon vacuum alignments  of the model presented in this paper are suggestive of geometry, and perhaps underlying crystalline structures. Investigating these avenues are the aim of a future publication \cite{prrsx2019}. 

\section*{Acknowledgment}
We thank L. Everett for her careful reading of the manuscript and helpful discussions and suggestions. \rd{We also thank the referee for their valuable suggestions and comments.} M.J.P. would like to thank the Departament de F\'{i}sica T\`{e}orica at the Universitat de Val\`{e}ncia for their hospitality during the preparation of this work. A.S. would like to acknowledge partial support from CONACYT Projects No. CB-2015-01/257655 (M\'exico) and No. CB-2017-2018/A1-S-39470 (M\'exico). M.H.R., P.R., and B.X. acknowledge partial support from the U.S. Department of Energy under Award No. DE-SC0010296.
\clearpage

\appendix

\section{$\mc T_{13}$ Group Theory} \label{app:T13}
The group $\mc T_{13}$ is second in the series $\mc T_n$, after its better-known sibling $\mc T_7$. In this appendix, we list the Kronecker products and Clebsch-Gordan coefficients of $\mc T_{13}$. For further details, see \cite{ishimori2012introduction}.

\subsection{Kronecker Products}
\begin{align*}
\g{1'} \otimes \g{1'} &= \gb{1}', \quad \g{1'} \otimes \gb{1}' = \g{1}\\[1pt]
\g{1'} \otimes \g{3}_i &= \g{3}_i, \quad \gb{1}' \otimes \g{3}_i = \g{3}_i\\
\g{3}_1\otimes\g{3}_1&=\gb{3}_1\oplus\gb{3}_1\oplus\g{3}_2\\[1pt]
\g{3}_2\otimes\g{3}_2&=\gb{3}_2\oplus\gb{3}_1\oplus\gb{3}_2\\[1pt]
\g{3}_1\otimes\gb{3}_1&=\g{1}\oplus\g{1}'\oplus\gb{1}'\oplus\g{3}_2\oplus\gb{3}_2\\[1pt]
\g{3}_2\otimes\gb{3}_2&=\g{1}\oplus\g{1}'\oplus\gb{1}'\oplus\g{3}_1\oplus\gb{3}_1\\[1pt]
\g{3}_1\otimes\g{3}_2&=\gb{3}_2\oplus\g{3}_1\oplus\g{3}_2\\[1pt]
\g{3}_1\otimes\gb{3}_2&=\gb{3}_2\oplus\g{3}_1\oplus\gb{3}_1\\[1pt]
\g{3}_2\otimes\gb{3}_1&=\g{3}_2\oplus\g{3}_1\oplus\gb{3}_1
\end{align*}

\subsection{Clebsch-Gordan Coefficients}
\begin{align*}
\matc{
\ket{1}\\
\ket{2}\\
\ket{3}
}_{\g{3}_1}
\otimes
\matc{
\ket{1'}\\
\ket{2'}\\
\ket{3'}
}_{\g{3}_1} &= 
\matc{
\ket{1} \ket{1'}\\
\ket{2} \ket{2'}\\
\ket{3} \ket{3'}
}_{\g{3}_2} \oplus
\matc{
\ket{2} \ket{3'}\\
\ket{3} \ket{1'}\\
\ket{1} \ket{2'}
}_{\gb{3}_1} \oplus
\matc{
\ket{3} \ket{2'}\\
\ket{1} \ket{3'}\\
\ket{2} \ket{1'}
}_{\gb{3}_1} 
\\[4pt]
\matc{
\ket{1}\\
\ket{2}\\
\ket{3}
}_{\g{3}_2}
\otimes
\matc{
\ket{1'}\\
\ket{2'}\\
\ket{3'}
}_{\g{3}_2} &= 
\matc{
\ket{2} \ket{2'}\\
\ket{3} \ket{3'}\\
\ket{1} \ket{1'}
}_{\gb{3}_1} \oplus
\matc{
\ket{2} \ket{3'}\\
\ket{3} \ket{1'}\\
\ket{1} \ket{2'}
}_{\gb{3}_2} \oplus
\matc{
\ket{3} \ket{2'}\\
\ket{1} \ket{3'}\\
\ket{2} \ket{1'}
}_{\gb{3}_2}
\\[4 pt]
\matc{
\ket{1}\\
\ket{2}\\
\ket{3}
}_{\g{3}_1}
\otimes
\matc{
\ket{1'}\\
\ket{2'}\\
\ket{3'}
}_{\g{3}_2} &= 
\matc{
\ket{3} \ket{3'}\\
\ket{1} \ket{1'}\\
\ket{2} \ket{2'}
}_{\g{3}_1} \oplus
\matc{
\ket{3} \ket{1'}\\
\ket{1} \ket{2'}\\
\ket{2} \ket{3'}
}_{\gb{3}_2} \oplus
\matc{
\ket{3} \ket{2'}\\
\ket{1} \ket{3'}\\
\ket{2} \ket{1'}
}_{\g{3}_2}
\\[4pt]
\matc{
\ket{1}\\
\ket{2}\\
\ket{3}
}_{\g{3}_1}
\otimes
\matc{
\ket{1'}\\
\ket{2'}\\
\ket{3'}
}_{\gb{3}_2} &= 
\matc{
\ket{1} \ket{1'}\\
\ket{2} \ket{2'}\\
\ket{3} \ket{3'}
}_{\gb{3}_1} \oplus
\matc{
\ket{2} \ket{3'}\\
\ket{3} \ket{1'}\\
\ket{1} \ket{2'}
}_{\gb{3}_2} \oplus
\matc{
\ket{2} \ket{1'}\\
\ket{3} \ket{2'}\\
\ket{1} \ket{3'}
}_{\g{3}_1}
\end{align*}
\begin{align*}
\matc{
\ket{1}\\
\ket{2}\\
\ket{3}
}_{\g{3}_2}
\otimes
\matc{
\ket{1'}\\
\ket{2'}\\
\ket{3'}
}_{\gb{3}_1} &= 
\matc{
\ket{1} \ket{1'}\\
\ket{2} \ket{2'}\\
\ket{3} \ket{3'}
}_{\g{3}_1} \oplus
\matc{
\ket{1} \ket{2'}\\
\ket{2} \ket{3'}\\
\ket{3} \ket{1'}
}_{\gb{3}_1} \oplus
\matc{
\ket{3} \ket{2'}\\
\ket{1} \ket{3'}\\
\ket{2} \ket{1'}
}_{\g{3}_2}
\\[4 pt]
\matc{
\ket{1}\\
\ket{2}\\
\ket{3}
}_{\g{3}_1}
\otimes
\matc{
\ket{1'}\\
\ket{2'}\\
\ket{3'}
}_{\gb{3}_1} &= 
\matc{
\ket{1} \ket{2'}\\
\ket{2} \ket{3'}\\
\ket{3} \ket{1'}
}_{\gb{3}_2} \oplus
\matc{
\ket{2} \ket{1'}\\
\ket{3} \ket{2'}\\
\ket{1} \ket{3'}
}_{\g{3}_2} \\
&\oplus  (\ket{1}\ket{1'}+\ket{2}\ket{2'}+\ket{3}\ket{3'})_\g{1}\\
&\oplus  (\ket{1}\ket{1'}+\omega \ket{2}\ket{2'}+ \omega^2 \ket{3}\ket{3'})_\g{1'}\\
&\oplus (\ket{1}\ket{1'}+\omega^2 \ket{2}\ket{2'}+ \omega \ket{3}\ket{3'})_{\gb{1}'}
\\[4 pt]
\matc{
\ket{1}\\
\ket{2}\\
\ket{3}
}_{\g{3}_2}
\otimes
\matc{
\ket{1'}\\
\ket{2'}\\
\ket{3'}
}_{\gb{3}_2} &= 
\matc{
\ket{2} \ket{3'}\\
\ket{3} \ket{1'}\\
\ket{1} \ket{2'}
}_{\g{3}_1} \oplus
\matc{
\ket{3} \ket{2'}\\
\ket{1} \ket{3'}\\
\ket{2} \ket{1'}
}_{\gb{3}_1} \\
&\oplus  (\ket{1}\ket{1'}+\ket{2}\ket{2'}+\ket{3}\ket{3'})_\g{1}\\
&\oplus  (\ket{1}\ket{1'}+\omega \ket{2}\ket{2'}+ \omega^2 \ket{3}\ket{3'})_\g{1'}\\
&\oplus (\ket{1}\ket{1'}+\omega^2 \ket{2}\ket{2'}+ \omega \ket{3}\ket{3'})_{\gb{1}'}
\\[4 pt]
\left(\ket{1}\right)_{\g{1}'} \otimes
\matc{
\ket{1'}\\
\ket{2'}\\
\ket{3'}
}_{\g{3}_i} &= 
\matc{
\ket{1}\ket{1'}\\
\omega \ket{1}\ket{2'}\\
\omega^2 \ket{1}\ket{3'}
}_{\g{3}_i}
\\[4 pt]
\left(\ket{1}\right)_{\gb{1}'} \otimes
\matc{
\ket{1'}\\
\ket{2'}\\
\ket{3'}
}_{\g{3}_i} &= 
\matc{
\ket{1}\ket{1'}\\
\omega^2 \ket{1}\ket{2'}\\
\omega \ket{1}\ket{3'}
}_{\g{3}_i}, \qquad \omega^3 = 1
\end{align*}

\newpage

\section{Enlarging the $SU(5)\times \mc T_{13}$ Symmetry} \label{app:shape}
The $SU(5) \times \mc T_{13}$ symmetry allows the following tree-level terms which have unwanted contributions to $Y^{(-\frac{1}{3})}$:
\begin{align}
&\overline{\Delta}\Delta\varphi^{(5)},\quad \overline{\Sigma} \Sigma\varphi^{(1),(2),(4)},\quad \overline{\Sigma} \Sigma{\varphi^{(6)}},\quad F\Delta\varphi^{(3)},\quad F\Delta{\varphi^{(6)}},\quad F \overline{\Sigma} H, \nonumber \\
&\overline{\Delta}\Delta\varphi^{(5)*}, \quad \overline{\Sigma} \Sigma(\varphi^{(1),(2),(4)})^*, \quad \overline{\Sigma} \Sigma{\varphi^{(6)*}}, \quad F \Delta (\varphi^{(1),(2),(4),(5)})^*,  \nonumber\\
& \Sigma T\varphi^{(1),(2),(3),(4),(5)},\quad T \overline{\Delta} H_{\overline{\g{45}}}, \quad F \Delta {\varphi^{(6)*}}, \quad \Sigma T (\varphi^{(1),(2),(3),(4),(5),(6)})^*. 
\end{align}
Suppose there is an Abelian $\mathcal{Z}_n$ symmetry whose purpose is to prohibit these terms.

We use $[~\cdot~]$ to denote the $\mathcal{Z}_n$ charges of the respective fields. For simplicity, we assume that $[F] = [T] = 0$.

From Eq.~\eqref{full_lagrang}, $[\mathcal{L}] = 0$. Setting the $\mathcal{Z}_{13}$  charge of each term equal to zero, we derive
\begin{align}
	\label{shape1}&[\varphi^{(1)}] = [\varphi^{(2)}] = [\varphi^{(4)}] = [\varphi^{(5)}] = [\overline{\Delta}] = k,\\
	\label{shape2}&[\varphi^{(3)}] = 0,\\
	\label{shape3}&[{\varphi^{(6)}}] = [\overline{\Sigma}] = k',\\
	\label{shape4}&[\varphi^{(1)*}] = [\varphi^{(2)*}] = [\varphi^{(4)*}] = [\varphi^{(5)*}] = [\Delta ] = [H_{\gb{5}}] = n-k,\\
	\label{shape5}&[{\varphi^{(6)*}}] = [\Sigma] = [H_{\overline{\g{45}}}] = n-k' ,
\end{align}
where $0 \leq k, k' \leq n$.

Now, prohibiting the unwanted terms listed above requires that their $\mathcal{Z}_n$ charges are nonzero, giving,
\begin{align}
	k \neq 0, \\
	k' \neq 0, \\
	2k \neq 0, \\
	2k' \neq 0, \\
	k \neq k', \\
	k + k' \neq 0.
\end{align}
The lowest triplet $\{k, k', n\}$ that satisfies these constrains is $\{1, 2, 5\}$. 

\noindent Then, Eqs.~\eqref{shape1}-\eqref{shape5} give the $\mathcal{Z}_{5}$ charges of the fields in the model.

\blue{Note that there are no non-Abelian groups of order equal to or smaller than five, hence $\mc Z_{5}$ is the \emph{smallest} symmetry that prohibits the unwanted terms.}

\newpage

\bibliography{frobenius}
\bibliographystyle{apsrev4-1}

\end{document}